# High Performance, Continuously Tunable Microwave Filters using MEMS Devices with Very Large, Controlled, Out-of-Plane Actuation

Jackson Chang, Michael Holyoak, George Kannell, Marc Beacken,
Matthias Imboden and David J. Bishop

***Abstract*— Software defined radios (SDR) in the microwave X and K bands offer the promise of low cost, programmable operation with real-time frequency agility. However, the real world in which such radios operate requires them to be able to detect nanowatt signals in the vicinity of 100 kW transmitters. This imposes the need for selective RF filters on the front end of the receiver to block the large, out of band RF signals so that the finite dynamic range of the SDR is not overwhelmed and the desired nanowatt signals can be detected and digitally processed. This is currently typically done with a number of narrow band filters that are switched in and out under program control. What is needed is a small, fast, wide tuning range, high *Q*, low loss filter that can continuously tune over large regions of the microwave spectrum. In this paper we show how extreme throw MEMS actuators can be used to build such filters operating up to 15 GHz and beyond. The key enabling attribute of our MEMS actuators is that they have large, controllable, out-of-plane actuation ranges of a millimeter or more. In a capacitance-post loaded cavity filter geometry, this gives sufficient precisely controllable motion to produce widely tunable devices in the 4-15 GHz regime.**

## I. INTRODUCTION

As contention for the electromagnetic spectrum continues to intensify, a number of strategies are being developed to cope. These include spread spectrum [1], ultrawideband [2] and cognitive radios [3]. These all require some version of software defined radio (SDR) [4]. A typical SDR architecture is a Direct-Conversion Receiver [5] (DCR) with a filter, quadrature detector and analog-to-digital converters to digitize the detector I/Q outputs for subsequent digital processing. The filter is needed because of the challenge of detecting nanowatt signals in the presence of powerful out of band transmissions and the finite dynamic range of the SDR. In this paper we discuss using novel MEMS devices in a capacitance-post loaded cavity filter geometry [6]. We believe such filters can meet the considerable challenges of being low

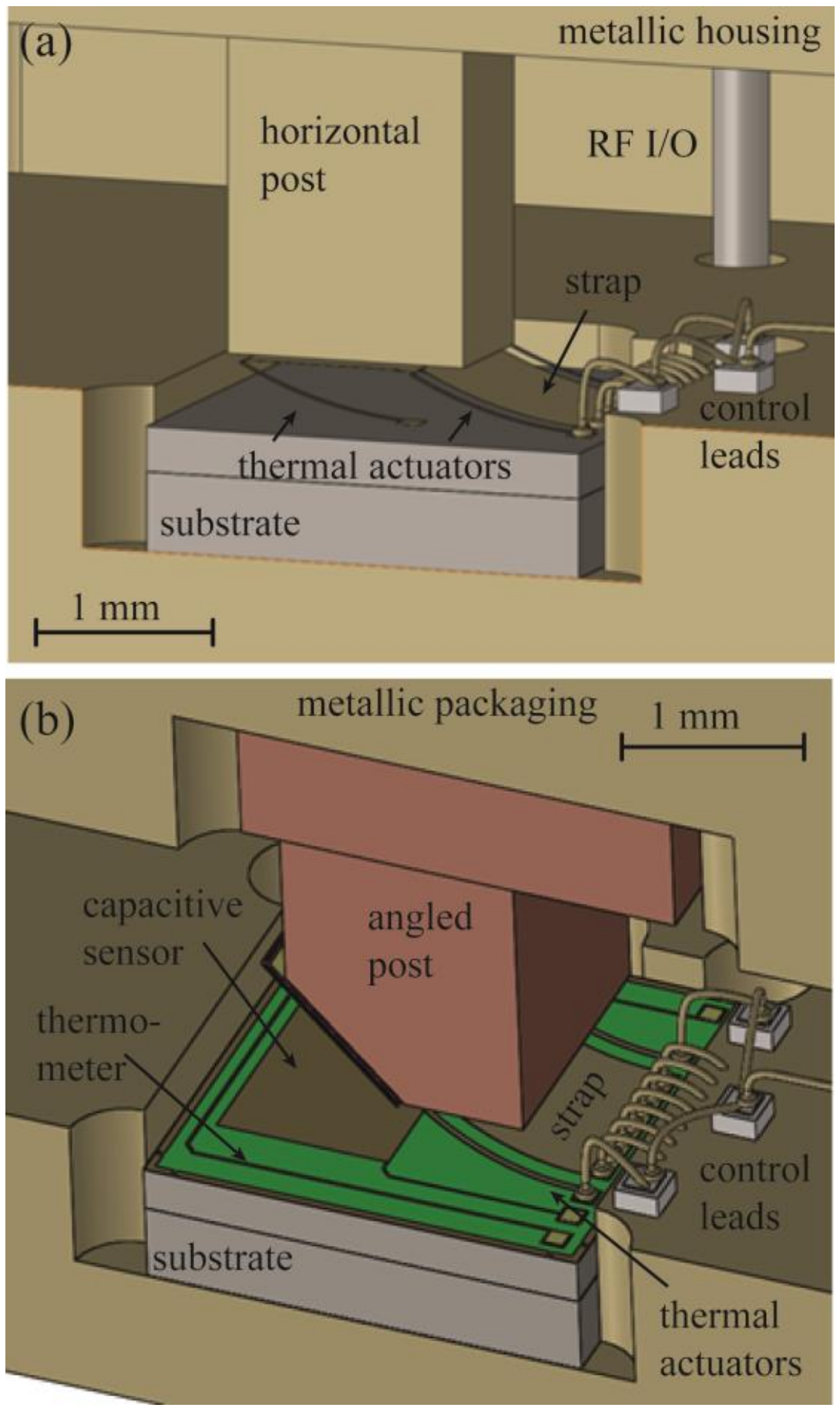

Figure 1. Shown is the conceptual layout of our device. A MEMS device moves a large plate relative to a post and tunes the capacitance causing a shift in the filter frequency. In the device in a), the plate remains parallel to the substrate and in b) the plate angle changes from being parallel to the substrate to parallel to the face of the machined post.

This work was supported in part by Boston University.
J. Chang is at Boston University, Electrical & Computer Engineering, Boston MA 02215 USA (e-mail: jchang16@bu.edu)
M. Holyoak is at LGS, Florham Park, NJ, USA (e-mail: mholyoak@lgsinnovations.com)
G. Kannell is at LGS, Florham Park, NJ, USA (e-mail: gkk@lgsinnovations.com)
M. Beacken is at LGS, Florham Park, NJ, USA (e-mail: marc@lgsinnovations.com)
Matthias Imboden is at École Polytechnique Fédérale de Lausanne (EPFL), STI-IMT LMTS, Neuchâtel, Switzerland
D.J. Bishop at Boston University, Electrical & Computer Engineering,

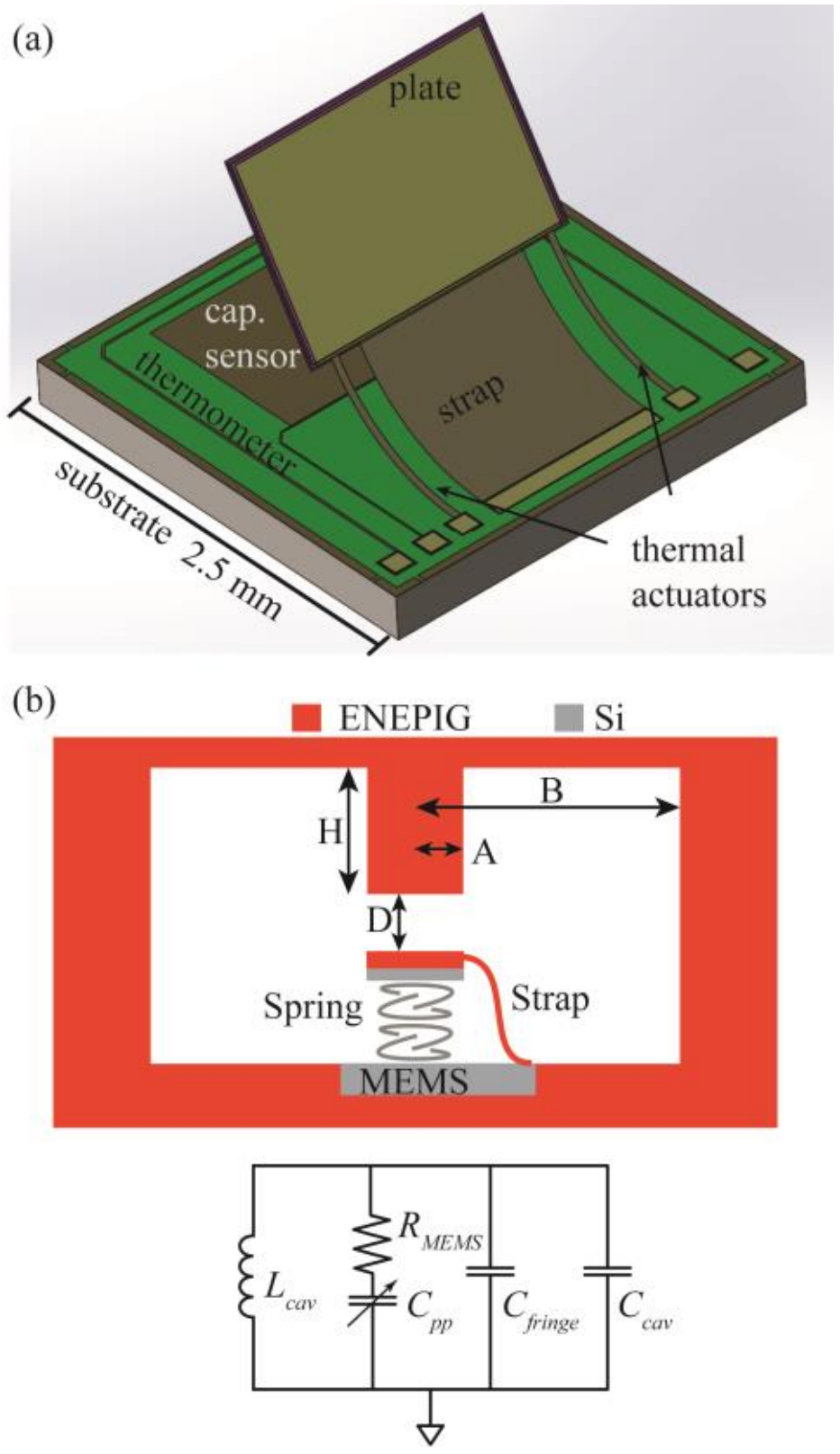

Figure 2. a) Shown is the concept for our MEMS device. There is a large, gold-coated MEMS plate which moves under the control of two bimorphs. Between the two bimorphs is a large, gold strap that reduces the RF insertion loss at the frequency of operation. There is a large fixed capacitor plate that is used to sense the position of the movable MEMS plate and a polysilicon thermometer for thermal sensing of the ambient temperature. (Alterantively the capaciticve sensing can also be performed using a guarded capacitor beneath the thermal actuators) b) Shown is a schematic of the filter. The crical dimensions are indeicated and an equivalent circuit model that defines the resoancne of the *LC* circuit (with losses) are illustrated.

loss, and having a wide tuning range with a small SWaP (size, weight and power).

MEMS-enabled active filters can work in either digital [7], [8] or analog modes [9]–[11]. In the digital mode, one can have switches for switching static filters in and out or digital varactors. Ref. [7] discusses a 4 bit, digital varactor approach. Here, we discuss an analog, continuously tunable MEMS microwave filter in a capacitance-post loaded cavity geometry. One major challenge in using MEMS devices in this application is that the typical maximum range of motion for MEMS actuators is ~10-100 microns, requiring careful engineering for C, X and K band operation. In this paper we describe a novel approach using MEMS devices with very large, controlled out-of-plane actuation of up to one millimeter and beyond.

A challenge with analog MEMS filters is the requirement for positional stability of the MEMS plate. The digital approaches discussed above avoided this problem. Here, we leverage the ability of MEMS device to be a “system-of-systems” so that we can integrate capacitance sensing elements into our MEMS die and use the information they provide for closed-loop control of the plate position. We can also integrate on-chip thermometers for correcting thermally-induced effects. Such MEMS devices have allowed us to build low RF loss, large tuning range X and K band microwave filters with a very small footprint. We believe these out-of-plane MEMS actuators may enable a new class of high performance microwave devices and systems.

Figure 1 shows the concept for our approach. A MEMS die is located inside a metallic housing with a post. In Figure 1, two types of posts are shown: in a) there is a post whose bottom surface makes up one plate of a tunable capacitor with the MEMS device being the second plate. In this configuration, the MEMS plate must move in a manner such that it is always parallel to the die substrate. In panel b) the post has an angled side, usually either 45 degrees or 60 degrees, and the MEMS plate rotates from being parallel to the substrate to being parallel to this angled edge. In essence, our device is an analog MEMS varactor in a capacitance post geometry. In such a filter, the inductance is provided by the annular cavity of the post and sidewalls.

Figure 2 a) shows a schematic of the MEMS 2.5x2.5 $mm^2$ MEMS die and b) illustrates a schematic for our filter. It uses a geometry similar to those in references [9]–[11]. In this example, the plate rotates and would be used in a housing such as shown in Figure 1 b). Figure 3 shows SEM images of the implementation of two versions of our MEMS devices. In a) of Figure 3 is a MEMS plate that remains parallel to the substrate during actuation and in b) is shown a device where the plate rotates, as in the drawing in Figure 2 a). As will be discussed later, the devices shown in Figure 2 and 3 solve a number of the design challenges of using MEMS devices for this application. The thermally driven bimorph actuators are capable of providing an extended actuation range for the MEMS plate, thereby providing a large RF tuning range. As discussed above, using a MEMS plate in an analog mode requires care in controlling its position precisely. The reason for the use of MEMS devices in a digital mode is that this eases the control requirements. We do not have this luxury and instead use a separate control circuit which includes sensing for feed-back and feed-forward control.

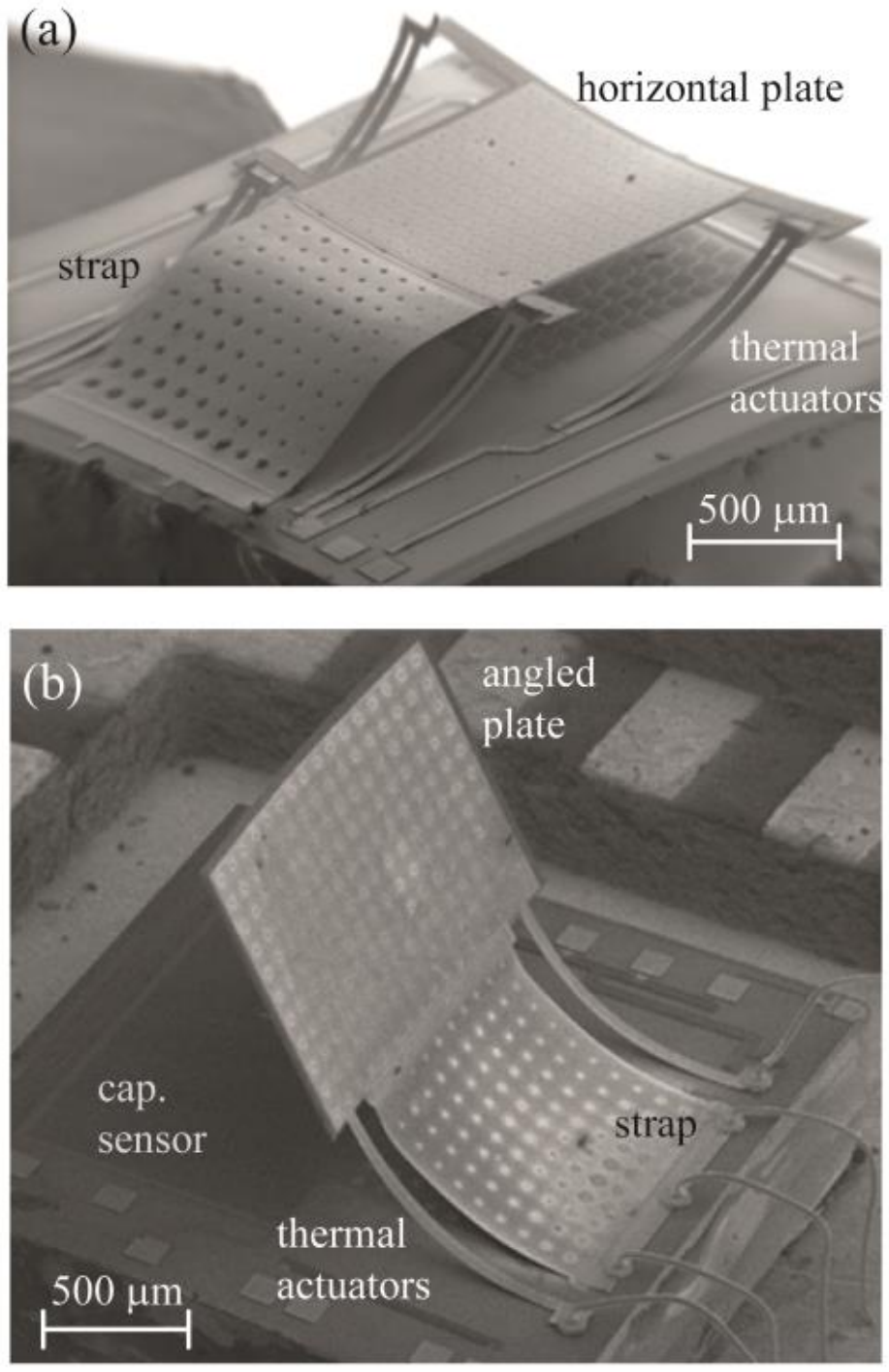

Figure 3. Shown are two examples of MEMS devices we have fabricated for use in our microwave filters. a) is a device where the movable plate remains parallel to the substrate during actuation. b) showns a plate that rotates from being parallel to the substrate to being parallel to an angled post during actuation. The devices shown in a) and b) of this figure would be used in the a) and b) housings shown in Figure 1, respectively.

MEMS devices are a "system of systems" enabling us to place additional elements onto the die for control functions. As shown in Figure 2, we can place a fixed capacitance plate, below the movable plate and/or below the thermal bimorphs, and use a low frequency capacitance measurement technique to sense (and control) the movable plate's position. Because these are thermal bimorphs, large changes in the ambient temperature will cause the position of the MEMS plate to vary. To allow us to measure and correct for this, we have also placed a polysilicon thermometer upon the substrate to monitor the ambient temperature. In subsequent sections this paper discusses the background of the application, the MEMS device fabrication and actuation, construction of the housing, the RF response, and finally the RF filter performance and modeling of the system.

Figure 2 b) shows the geometry of our filter. The filter is an *LC* resonator. The inductance is given by the post and the cavity, and the MEMS plate and post comprise the capacitor. Moving the MEMS plate relative to the post changes the capacitance and hence the resonance frequency. Within a lumped element approximation the following expressions describe the resonance frequency of the device, resulting from the capacitances and inductance [11]:

$$f_0 = \frac{1}{2\pi\sqrt{L_0(C_{fringe}+C_{cav}+C_{pp})}}, \tag{1}$$

and

$$L_0 = \frac{1}{2\pi}\mu_0 \ln\left(\frac{B}{A}\right), \tag{2}$$

$$C_{pp} = \varepsilon_0 \frac{A^2}{D}, \tag{3}$$

$$C_{fringe} = 2.8\varepsilon_0 A \ln\left(\frac{H}{D}\right), \tag{4}$$

$$C_{cav} = 2\pi\varepsilon_0 H \ln\left(\frac{B}{A}\right), \tag{5}$$

$$Tuning\ Ratio = \frac{f_{max}}{f_{min}} \sim \sqrt{\frac{D_{max}}{D_{min}}}, \tag{6}$$

where $f_0$ is the resonance frequency, $L_0$ is the inductance and $C_i$ the capacitances describing the device. The geometric parameters *A*, *B*, *D*, and *H* are described in Figure 2 b). $f_{max}$ and $f_{min}$ define the tuning ratio. The inductance is a result of the post/cavity geometry and the three capacitances can be described by: the plate/post capacitance-$C_{pp}$, the fringing capacitance between the plate and the cavity walls-$C_{fringe}$, and the cavity capacitance-$C_{cav}$. The resonant frequency of the *LC* circuit is given by the inductance and sum of the three capacitances. The equations present are oversimplified and are included to illustrate the concept and the importance of a large throw as it translates into a large running ratio. Rough (idealized) estimates indicate a tuning ratio of order 10:1 may be attainable, while as shown below, the experiments and more detailed numeric simulations result in a tuning ratio of ~3:1. This can be explained by the cavity capacitance and the imperfect throw of the device (both in terms of reaching the substrate and the closest approach which is limited by the alignment accuracy and plate flatness). The tuning ratio is the change in center frequency as one changes $C_{pp}$. In the limit where $C_{pp} >> C_{fringe}$, $C_{cav}$, the expression for the tuning range simplifies to $(D_{max}/D_{min})^{1/2}$, where *D* is the distance between the movable MEMS plate and the post. This expression shows why a large throw for the actuator is required for a large tuning range of the filter. For example, because of plate curvature, the typical closest approach ($D_{min}$) for our device is ~10 microns. With a throw of 1000 microns, a tuning range of 10 is possible. Our best case, experimentally observed tuning range is a factor of three. But with more careful engineering of the other capacitances, much larger tuning ranges are possible. This analysis shows why a large throw MEMS actuator is such an advantage for this type of filter.

## II. Background

RF receivers require a narrowband filter on their front end. This is due to the very high dynamic range over which they must operate from low power levels up to the levels of interfering signals, approaching 20 orders of magnitude in power (-120 dBm ->+80 dBm). For software defined radios (SDRs) that are frequency agile, this requires that the front end filter be frequency agile as well. In the early days of microwave technology, tunable filters were often trombone filters, a large, macro-mechanical device. Today, the standard implementation for frequency agile filters on the front end of SDRs are switched RF filter banks. However, these banks can be large, heavy, power hungry objects. Tunable filters may allow for much less SWaP than these banks if they can achieve the other required system specifications such as insertion loss, *Q*, tuning speed, power consumption, power handling, size and cost. Candidate technologies for tunable filters include III-IV varactors [12], ferroelectric materials [13], YIG ferrites [14] and RF MEMS [7]–[11]. RF MEMS devices are competitive with these other technologies in terms of the required systems specifications mentioned above and have the added potential benefit of being able to be integrated with the other electronic components of the radio in the same way the mechanical and electronic components for accelerometers are all integrated onto a single chip today. Thus a mechanical implementation of microwave filters may be the optimal technical solution, albeit a factor of a thousand times smaller in linear dimensions from the trombone tuners of five decades ago.

Some compelling MEMS microwave filters today are digital, using a bank of digital varactors [7]. These filters have 16 different states and provide near-continuous coverage over their tuning range but do so by having a relatively low filter *Q*. As the filter *Q* increases, a digital varactor approach will reduce the tuning range or increase the number of digital states in order to provide continuous coverage. More digital states require more real estate for the filter and SWaP will degrade. The advantage of a continuously tunable approach is that tuning range and *Q* can be separately optimized while still providing for continuous coverage over the RF range [11]. This is the approach we have adopted for the work described in this paper.

In general, such a filter pushes MEMS technology in directions it normally is not optimized for. Large tuning ranges require large displacements, a millimeter or more. Analog tuning requires good open loop stability along with some means for measurement and feedback/control of the plate position. Analog control also requires control over effects due to thermal drifts. MEMS do have advantages in terms of cost, SWaP, *Q*, power handling, tuning speed, insertion loss and linearity. In this paper we show how using thermal bimorphs, we can obtain very large throw actuators, a millimeter or more. We also show how with the additional post-processing, one can take a commercial MEMS foundry process (MEMSCAP PolyMUMPs [15]) and add free standing metallic straps that reduce RF insertion losses and keep *Q*'s high. By integrating on-chip capacitance measurement devices and thermometers, we can minimize the issues associated with analog control. At the end of the day, we believe the intrinsic advantages of analog MEMS microwave filters can outweigh their challenges and show here, prototypes of high performance RF filters.

## III. Device Design and Post Processing

The devices shown in Figure 3 are fabricated with the PolyMUMPs process [15]. The standard process was modified to substitute the carrier wafer with low resistivity (< 0.001 ohm-cm) Si wafers which are held at ground potential. The devices consist of three basic elements. Gold coated polysilicon cantilever beams form curved thermal actuators due to residual strain. In the case of the parallel plate device (Figure 1 and Figure 3a)). The connection between the plate and the four actuators is formed by a serpentine polysilicon spring. The strain in these springs (visible partly in Figure 4 b)) imposes the limiting factor on the maximum height reached (the mechanical restoring force due to the metal strap is weak by comparison). Applying a voltage causes Joule heating and straightens the thermal actuator resulting in out-of-plane plate actuation. A 1.5x1 or 1x1 $mm^2$ plate, which makes up one side of the tunable RF capacitor, is connected to the free end of the actuator. In order to ensure that the plate is flat, it is made up of two polysilicon layers encapsulating a silicon dioxide layer. Lastly, a free floating gold strap connects the plate to the grounded substrate in order to maximize the reflective area and to minimize the electrical resistance of the plate to ground (see Figure 2, Figure 3, and Figure 4).

The gold strap connections were made with additional post-production fabrication steps. Based on the skin depth of the PolyMUMPs gold layer and the intended RF frequencies, simulations indicate that in order to minimize RF losses the plate should have a 1 mm wide gold connection. However, fabricating such a connection without violating process design rules would result in a 1 mm wide gold/polysilicon beam that requires significantly more force to actuate than the actuators can provide, or drastically increase the thermal power required to produce the desired travel range. Omitting the polysilicon under the gold would resolve the rigidity issue and result in a gold strap but introduces fabrication issues. Because of the poor step coverage from the physical vapor deposited gold, the height mismatch between the plate and strap and between the strap and pad results in a physically disconnected and electrically open strap. To address this, additional post-fabrication processing is necessary. These steps are outlined in

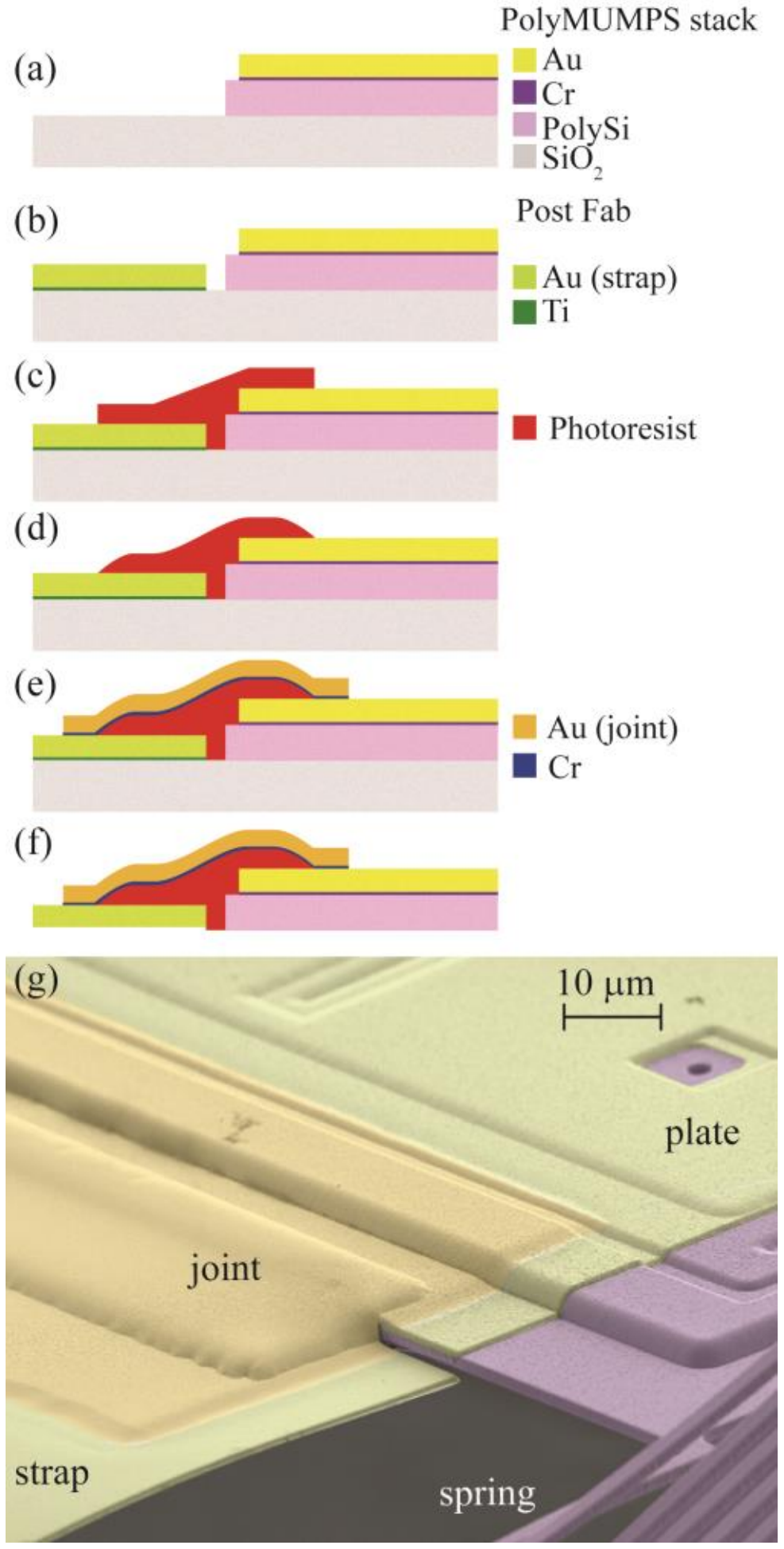

Figure 4. Shown is the post processing done to connect (mechanically and electrically) the large strap to the capacitor plate and the ground. a) As fabricated by the PolyMUMPs process. b) Depoition (Ti/Au) of the strap that electrically connect the plate to the substrate c) Patterned photoresist to form the mechanical joint connection d) Reflow of the photoresist e) Deposition of Cr/Au to create an electrical connection. f) HF release removes the sacrificial oxide and the Ti beneat the strap. g) SEM image of the resulting device.

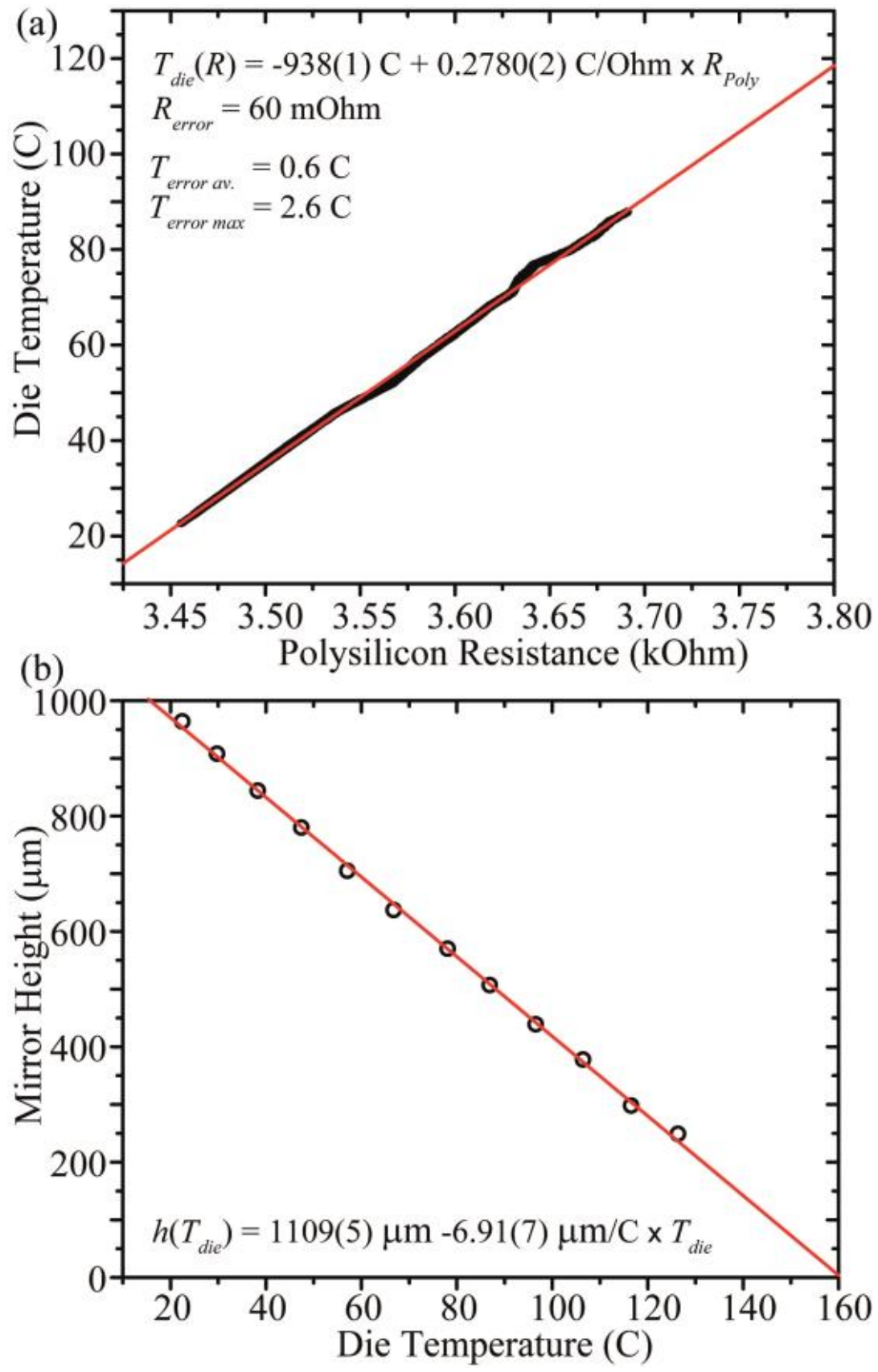

Figure 5. Thermal charactersitics. a) the on chip polysilicon thermometer can be used to monitor the temperature of the MEMS device. b) Temperature vs height plot illustrates the effecet of the die temperature on the height of the plate. These calibration measurements are needed for reliable operation of the devices, or show the need for environmental control.

Figure 4. After the PolyMUMPs processing is complete the gold strap (0.5 microns thick) is patterned direclty onto the sacrificial oxide layer and deposited using standard optical lithography and e-beam evaporation. Titanium is used for adhesion as it minimizes intrinsic strain and is fully removed by the final HF release step. For adhesion to the substrate and top plate a strip of photoresist is then patterned over each end of the strap. By heating the photoresist above its glass transition temperature, the photoresist reflows and its edges become rounded and serve as a smooth substrate for a subsequent evaporation of gold to connect the strap at either end. In this case chromium is used as an adhesion layer as it has a negligible etch rate in HF and will maintain a connection between the strap and the joint that survives the relae step.. The strip of photoresist serves as the mechanical joint while the gold from the second post fabrication evaporation serves as the electrical short between the plate and strap and between the strap and pad. These fabrication steps are illustrated in Figure 4. Typical layer thicknesses for the strap and adhesion layers are ~500 nm of gold and 15-20 nm of adhesion layer (Ti or Cr). These thicknesses are similar to the native PolyMUMPs Au layer. A detailed account fo the silicon, oxide and metal stack forming the MEMS is provided by [15]. The PolyMUMPs foundry process is very stable with device layer thickness stability better than 1 %. The minimum feature size is 2 µ and alignment between layers of the same order. Based on the data provided by MEMSCAP the greatest process variation is in the tensile strain of the metal (30% variation) and the compressive strain in the polysilicon layer 2 (18% variation). The spread in these strain levels are important for the curvature of the thermally actuated bimorph. A rapid thermal annealing step of the completed device ensures sufficient strain in the bimorphs for them to raise the plate to the post. With this set as the maximum height (smallest gap D) and the ability to actuate the bimorphs and lower the plate to the substrate (largest gap D) the full capacitance range, and hence tuning factor is ensured. (In parctive most of the range in freqeucny is achieved before the plate reaches the substrate is is seen below, making contact with the post however is critical to maximaize the tuning range).

On-die environmental sensors are designed to track thermal fluctuations that may impact the device performance. A polysilicon wire running along the substrate acts as a thermometer by utilizing its well-characterized coefficient of thermal resistance [16] (see Figure 5). Fixed polysilicon pads under the bimorph actuators and/or plate provide plate position sensing by forming a variable capacitor, whose capacitance is directly correlated to the plate position (see Figure 6). In the filter with the angled post, the MEMS plate does not remain parallel to the post surface throughout the travel range. The MEMS capacitive sensor is on the surface of the MEMS substrate and the MEMS plate is initially at an angle with respect to the sensor (and parallel to the plate surface). As power is applied to the thermal bimorphs, the angle of the MEMS plate with respect to the sensor decreases (as the plate rotates away from the post and towards the sensor on the substrate). This range of travel creates a near-linear relationship between capacitance and power applied. The AD7747 capacitance to digital converter chip is used to measure this capacitance, providing real-time sensing of the plate position. This, along with the thermometer data can be used to generate feedforward algorithms to more accurately control the position of the plate and consequently the filter frequency. Additional polysilicon pads placed in between the substrate and sensing pads are driven with an active shield signal to cancel any parasitic sensing capacitance to the substrate. The device is released in a timed wet HF bath tailored to ensure full release of the plate and thermal bimorphs but leave an incomplete etch of the oxide in between the sensing and active shield pads. After critical point drying, a 20-40 nm layer of $Al_2O_3$ is deposited by ALD to passivate the actuators. This ALD step has two functions. First, it means the MEMS plate does not electrically short to the packaging post, even when it comes in physical contact. Second, the ALD layer stabilizes the gold films of the bimorphs and makes them much more mechanically stable [17]. While the exact process for this increase in mechanical stability is not well understood, it is hypothesized that the $Al_2O_3$ coats the Au grain boundaries and keeps the grains from annealing which would relax the induced strain in the bimorph, causing drift.

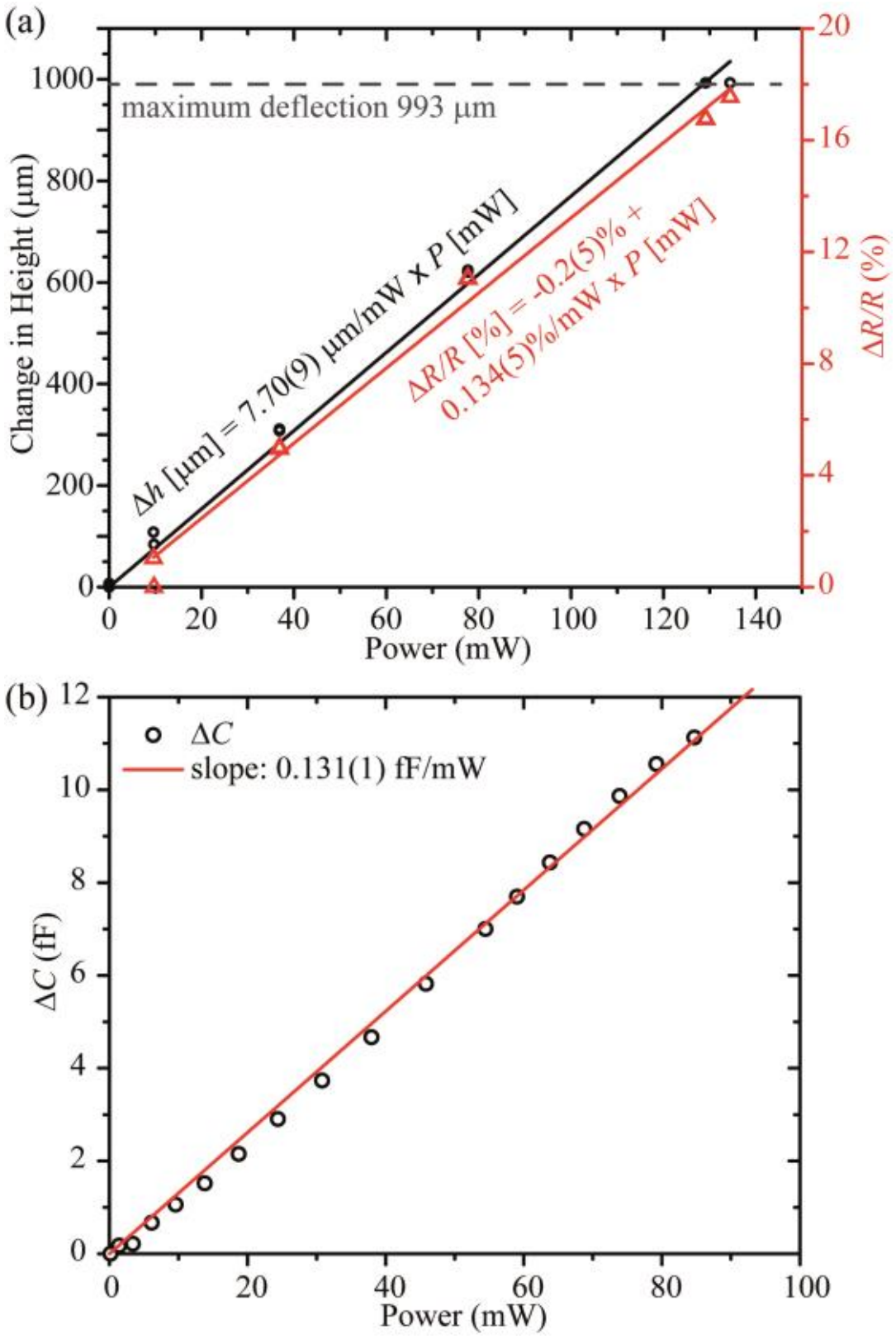

Figure 6. Actuation and Sensing. a) Power characterics for an angled plate device. The change in height is plotted against the total power applied to the thermal bimorphs indicating that a height change of 1 mm can be obtained by applying 140 mW of power. The change in resistance, due to heating can be used for sensing. Changes in the measured capacitance are shown in (b). A linear realation is found with a power dependence of 131 aF/mW, correspding to 58.8 mm/fF.

Table 1 shows data from an actuator similar to that shown in Figure 3, b). Note the large displacement which is key to the performance of our filter. The corresponding plots of the height and bimorph resistance is given in Figure 6. The height dependence on the power applied to the thermal actuators is plotted in Figure 6 a), showing a linear power height relation for a typical device. Figure 6 b) shows the linear change in capacitance with respect to the power applied to the plate. The power was determined from measureing the current for each applied voltage, all parameters are summarized in Table I. Consequently, a linear capacitance-height relationship can be extracted. The thermal actuators also have a temperature dependent resistivity which can also be used to determine the position of the plate once calibrated (a requirement for feed forward operation). Given that the plate height is driven by thermal actuators one would expect the package temperature to affect the plate height as well. Figure 5 a) shows resistance vs temperature measurements for the on chip polysilicon thermometer, illustrating the ability to track environmental changes in close proximity to the actuators and plate. Figure 5 b) plots the plate height as a function of package temperature. This information can be used with feed-forward algorithms to model the height of the plate without the capacitive feedback measurements. It also indicates that the package needs to operate in a controlled environment to ensure that the full actuation range is available. In general the feed-forward allows for imprived response time, however for high accuracy and stability feedback algorithms, with the input from capacitive sensing data will be required. A video sequence of the actuated device is included in the supplementary materials.

TABLE I
RESPONSE OF A BIMORPH ACTUATOR SIMILAR TO THAT SHOWN IN FIG. 3. NOTE THE LARGE AMPLITUDE OF DISPLACEMENT FOR THIS ACTUATOR.

| Voltage (V) | Current (mA) | Power (mW) | Resistance (Ohm) | Height (μm) |
|---|---|---|---|---|
| 0.0 | 0 | 0 | - | 1009 |
| 0.1 | 97 | 9.7 | 1.03 | 925 |
| 0.2 | 184 | 36.9 | 1.08 | 701 |
| 0.3 | 259 | 77.7 | 1.16 | 384 |
| 0.4 | 323 | 129.2 | 1.24 | 16 |
| 0.41 | 328 | 134.4 | 1.25 | 16 |
| 0.4 | 323 | 129.2 | 1.24 | 16 |
| 0.3 | 259 | 77.6 | 1.16 | 389 |
| 0.2 | 184 | 36.9 | 1.09 | 698 |
| 0.1 | 96 | 9.7 | 1.03 | 910 |
| 0.0 | 0 | 0 | - | 1002 |

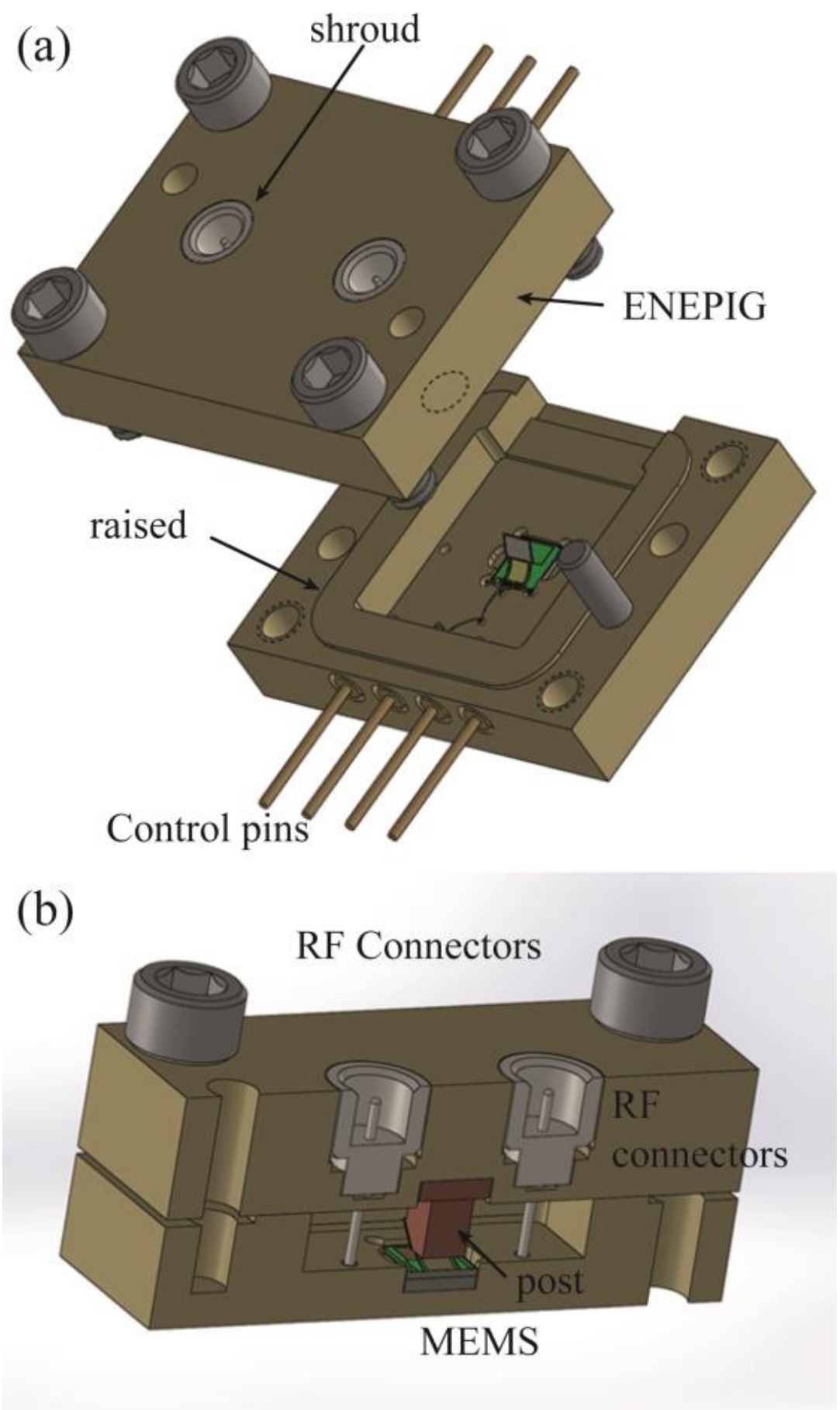

Figure 7. a) shows the exterior of version 1 of our RF filter package. The RF connectors (2) are located on the top half of the package. The MEMS device is located on the bottom half. The four low frequency leads on the bottom half of the package are used for the bimorph drive and the position and thermal sensing elements on the die. The interior of the packaging is shown in b).

## IV. Package Concept

Packaging the MEMS RF filter is a crucial part of our solution. While it is generally true for all MEMS devices, it is especially so for this one. Shown in Figure 7 is a diagram of our version 1 package. It comprises two halves. The top half contains the two RF connectors plus the post, either flat or angled (see Figure 1). The bottom half has a well into which the MEMS die is placed. The bottom half also has the low frequency leads for driving the bimorph and for making electrical contact to the capacitance position sensor and the thermal sensor. The housing is made from OFHC copper, plated with ENEPIG (Electroless Nickel Electroless Palladium Immersion Gold). In ENEPIG plating, the palladium acts as a barrier layer between the copper housing and the gold, avoiding the formation of intermetallics that reduce the gold conductivity and increase the RF losses. The coating is 1.2 microns thick. Alignment pins assure accurate positioning as the two halves are assembled. The RF and low frequency connectors are hermetic and a solder preform can be used between the two halves to ensure the hermeticity of the package. The packages can be assembled in a dry nitrogen atmosphere to ensure long-term reliability. If desired, other gases could be used to adjust the thermal response or environmental performance. For a reliable commercial product the hermetic seal is critical.

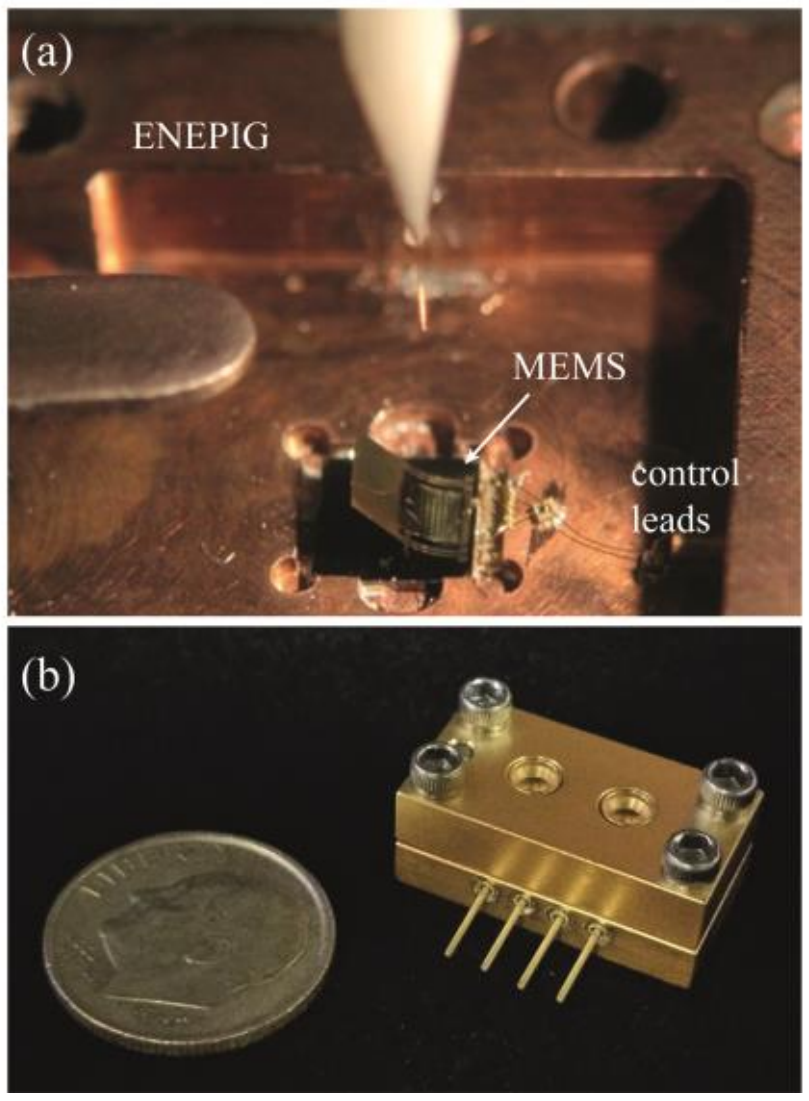

Figure 9. a) Shown is the inside of the package with a MEMS device mounted. b) Shown is the exterior of a version 2 package. A dime is shown for size reference.

Figure 9 a) shows an optical image of the bottom half of the package with a MEMS device in place. At the center top is the wire and capillary of the ball bonder used during assembly, giving the size scale. The lower panel in Figure 9 shows our version 2 package. The interior is similar to that shown in Figure 7 but extraneous metal has been removed to reduce the overall size of the package. A dime is shown for size comparison. In Figure 9 a) one can just barely see the wire bond connections between the low frequency feedthroughs and the MEMS die. These wirebonds are shown much more clearly in Figure 8 which shows SEM images of the inside of the package with a MEMS device mounted. Electrically, the RF feedthroughs on the top half of the package make good electrical connection to the lower housing. This is shown in detail in the lower panel of Figure 7. The MEMS plate is connected to a large gold strap and then that strap is connected to the housing by ~7 wire bonds. Figure 8 b)) shows a device with an on-chip thermometer. The images (a) and b)) show the wirebonds to these and their connections to the low frequency feedthroughs. Also shown in Figure 8 are filter capacitors which keep the RF signals from coupling to the low frequency leads. These capacitors are the dark, square objects to the right of the MEMS die in the lower image and are connected to the housing with conductive epoxy. There are no issues with parasitic inductances nor parasitic capacitance from the wire-bonds or single layer capacitors. The components' surface areas realtive to the cavity dimensions and the components' location in the cavity make their impact to the resonance frequency insignificant.

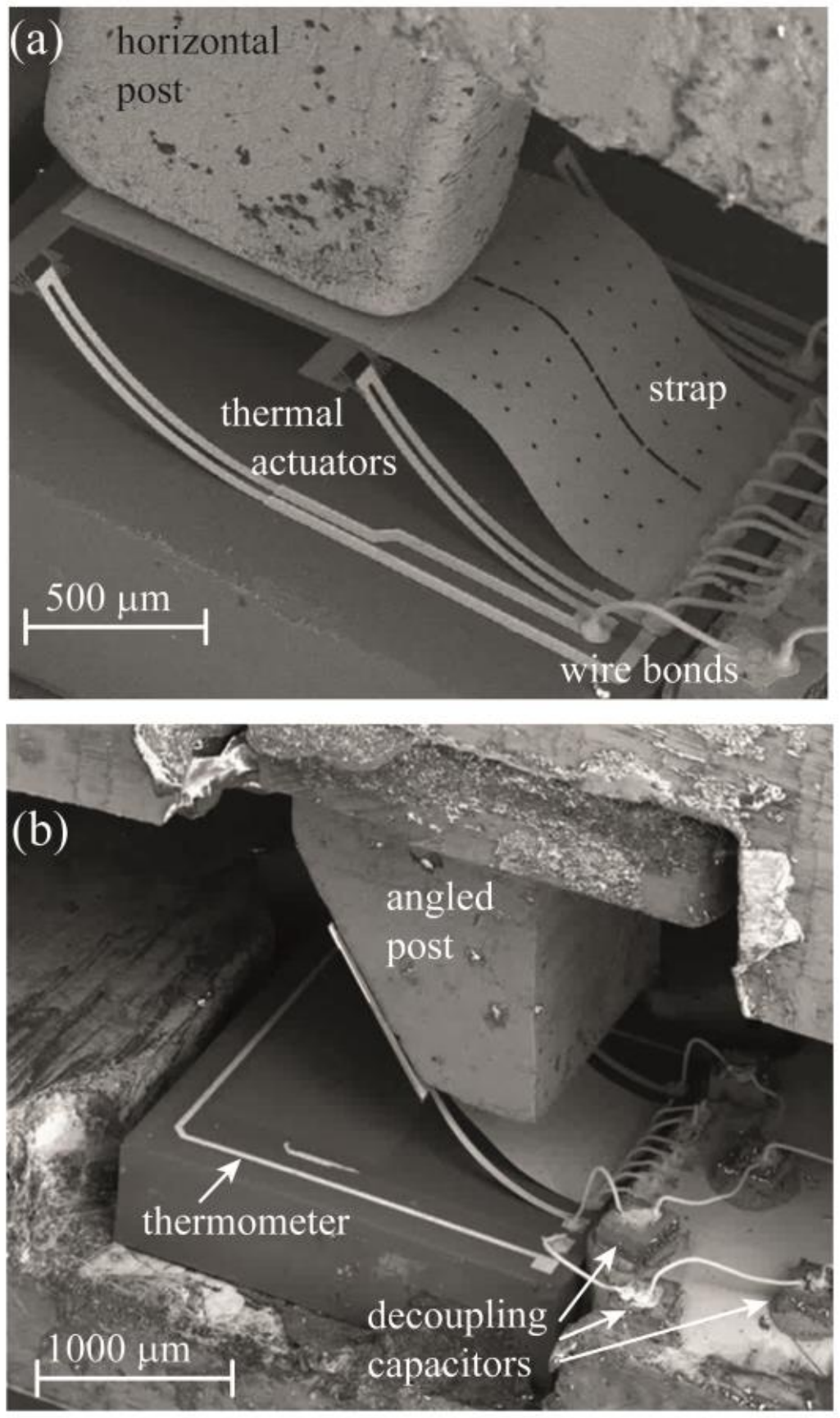

Figure 8. Shown is the interior of the package with a MEMS device mounted. In the upper panel, shown is a right-angled post and in the lower panel, an angled, 60 degree post.

## V. Data and Analysis

Figure 10, Figure 11, and Figure 13 show results from two port transmission measurements of our packaged sub-system. The measurements were done using a Keysight E8364B PNA

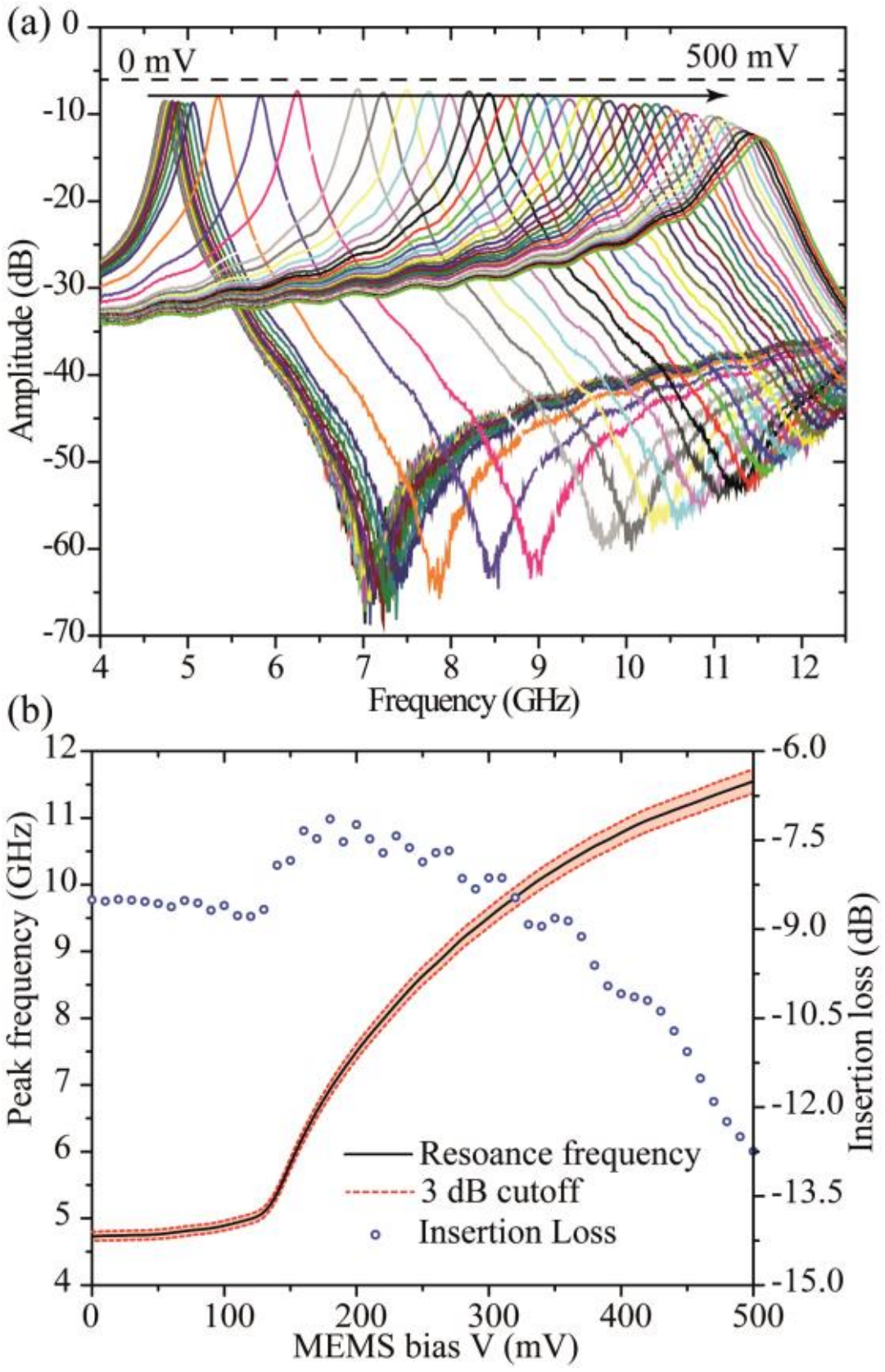

Figure 10. Shown are transmission results for our parallel plate (0°) device between 4 and 12 GHz. The bimorph voltages for each plate position are set from 0 to 500 mV in 10 mV increments. b) Filter center frequency and insertion loss as as function of the MEMS bia voltage.

Network analyzer. This unit allows measurements of up to 50 GHz to be performed. Measured quantities included both the transmission and return losses. A typical scan used 16,000 points with a measurement speed of ~100 microseconds/point for a complete scan lasting over several seconds. Typical power levels used were -12 dBm. We estimate errors in our measurements of the absolute transmission to be less than 0.5 dB. The MEMS is controlled using a voltage bias which is sweeped from 0 mV to 500 mV in 10 mV increments. The equivalent power can be inferred from Table I.

Figure 10 shows a typical transmission measurement between 4 and 12.5 GHz. Each trace is a scan for a particular voltage across the bimorph actuator corresponding to a particular plate height. The bimorph voltage is incremented from 0 mV to 500 mV in 10 mV steps, the height can be inferred from Table I. For each voltage (and plate position) a complete scan was taken between 4 and 12.5 GHz. This device used the package shown in Figure 8 a) with a 90 degree post. Several things are clear from this plot. The first is that the center frequency of the device can be tuned over a wide range (4.7 GHz to 11.4 GHz) for a tuning range exceeding a factor of two. The second point is that the $Q$ is relatively narrow (~100) and remains constant over the tuning range. The insertion loss is moderate (better than 10 dB) over most of the tuning range. As will be seen in the RF modeling section, circuit loading results in lower quality factor and higher insertion loss when the plate is in close proximity to the post. For the parallel MEMS-post device however, RF loss is MEMS dominated, resulting in high insertion loss over all frequencies. This is due primarily to the lossy polysilicon springs connecting the bimorphs to the plate, which are not included in the angled MEMS device.

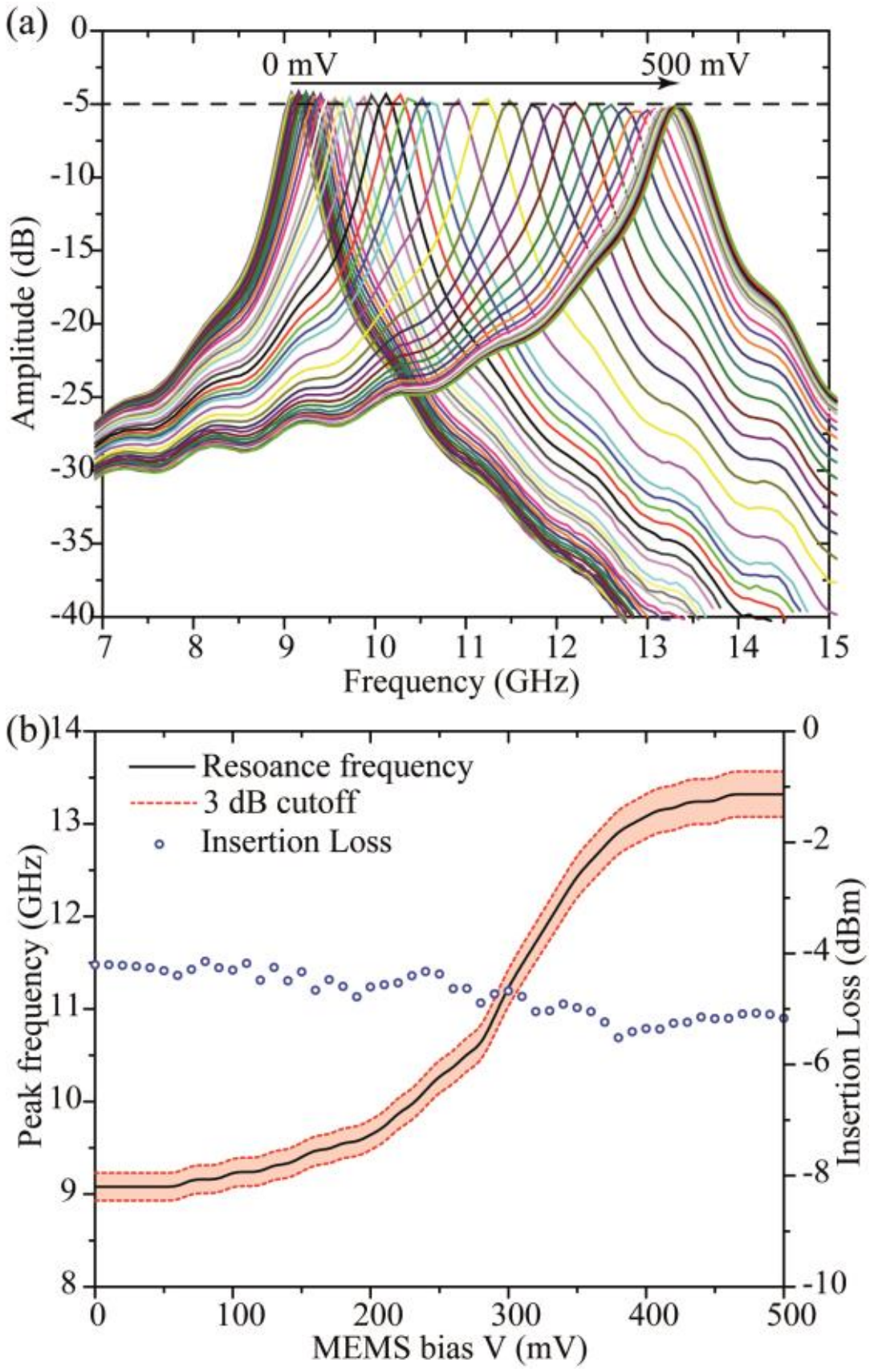

Figure 11. RF filter with one MEMS and a 45˚ post. a) Shown is the amplitude response ($S_{21}$) of an optimized device operating between 7 and 14 GHz. Note the low overall losses, approximately 5 dB. The bimorph voltages are incremeneted by 10 mV from 0 to 500 mV. b) Filter center frequency and insertion loss as as function of the MEMS bia voltage.

Figure 11 shows an optimized device with at 45 degree post. It uses a single MEMS plate (Figure 8 b)) and the scans range from 7 to 15 GHz. The center frequency of the filter ranges from 9.2 GHz to 12.5 GHz. Note the much lower losses, of order 5 dB over the tuning range. Based on simulations (discussed below) we believe losses less than 2 dB are possible. The losses are attributed to imperfect alignment of the MEMS and post as the assembly procedure introduces variations from the ideal. Figure 13 shows data from a device with an angled plate and a 60 degree post, operating from 4 to 18 GHz with losses below 3 dB over part of its operating range. This device has a tuning range ~3 with $f$ ranging from 5.5 to 15 GHz. As the tuning voltage decreases, the cavity resonator becomes increasingly loaded by the capacitance of the MEMS plate and fixed post. This in turn with the skin depth of the MEMS Au layer decreases the resonant frequency amplitude and quality factor. Moreover, impedance is affected and the resulting return loss is higher. EM simulation confirms this performance across tuning range and is illustrated in section VI.

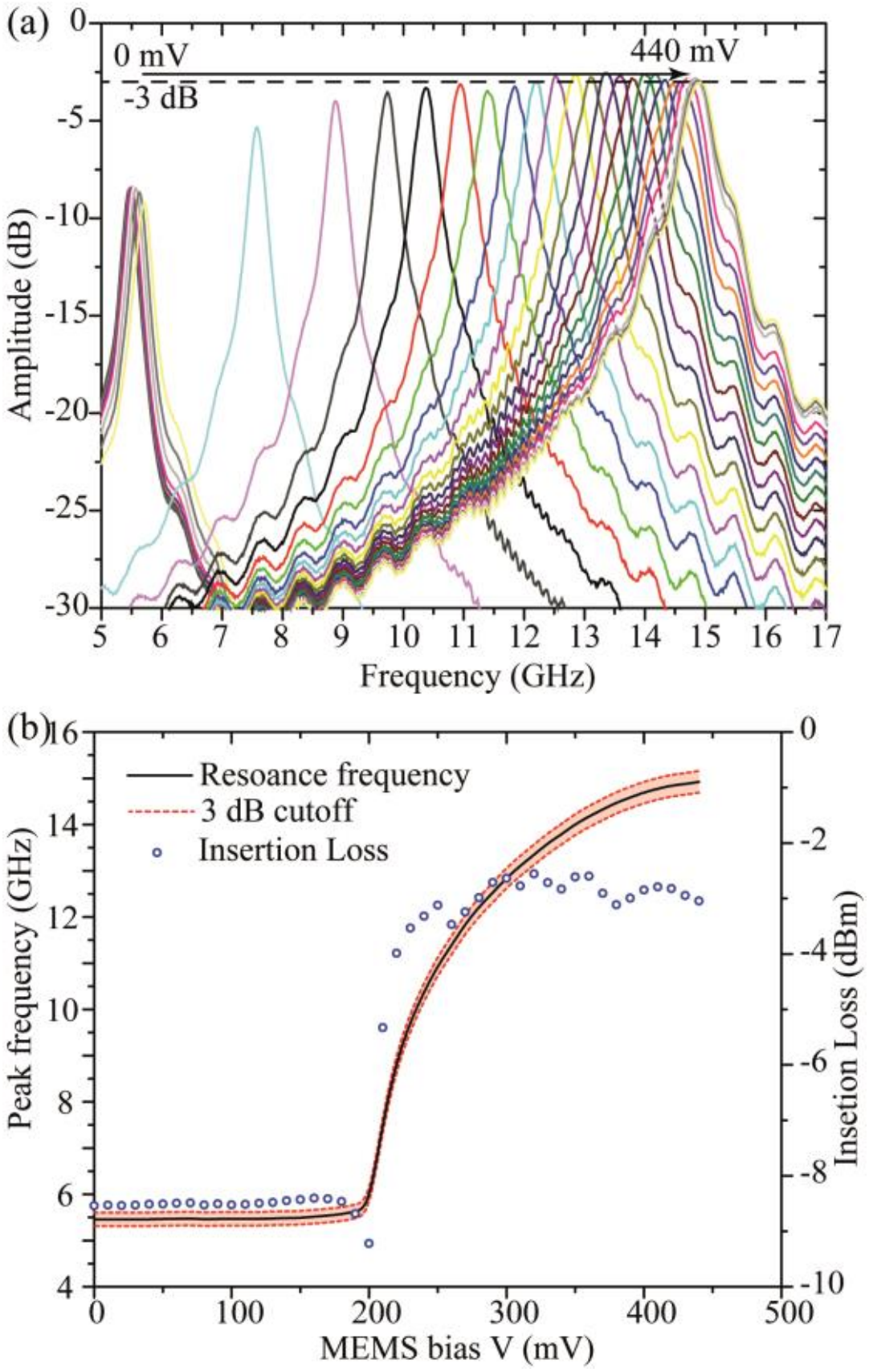

Figure 13. Shown is the operation of our angled plate device with a 60° post over a 5-17 GHz frequency span. Note that over parts of the operating range, the insertion losses are as low as 2.5 dB. The voltage is increased in 10 mV increments from 0 to 440 mV. b) Filter center frequency and insertion loss as as function of the MEMS bia voltage.

An important point about our device is that there is no intrinsic cutoff frequency. Our device is essentially a metal plate (gold) moving in a gold plated metal housing. There is no bandgap or other intrinsic frequency scale for its operation. By careful design and care with features in the package that create loss, we believe operation at much higher frequencies is possible. This means that optimizing one MEMS device results in a technology that can work over a very wide frequency range, determined by the package design. In the results shown here, we have one MEMS plate moving relative to a fixed metallic post. However as shown in Figure 14, one can also build devices with two MEMS plates and actuators. This doubles the maximum value for *D* and correspondingly increases the tuning range.

## VI. RF Modeling

RF modeling has been an important tool in the development of our MEMS-enabled RF filter system. While lumped element models are helpful as rough intuitive guides, true understanding and optimization of our device requires full EM modeling. The primary system we have used for this is the Keysight EMPro 3D Simulation Software. Without such modeling we would have spent considerably longer in developing optimal packaging solutions. Using the simulation package, we are able to take 3D models such as shown in Figure 7 and calculate the transmission and reflection losses. We can use digital design to let us optimize our MEMS die and package combination and then construct systems that closely mirror what we have simulated. Shown in Figure 12 is the simulated performance of the MEMS filter. The different plots depict the S-parameters of the filter as the MEMS plate angle with respect to the substrate is varied from 5 degrees to 40 degrees in 5 degree steps. A plot is also included that sets the angle to 42 degrees, which is representative of the MEMS plate being parallel with a 45-degree angled post as shown similarly in Figure 1; the separation distance between the MEMS plate and angled post is approximately 20 μm. S21 and S11 plots represent insertion losses and return losses, respectively. Some noteworthy points are that (1) the simulated filter demonstrates comparable performance with the measured responses as shown in Figure 13, although measurements do indicate higher insertion losses across the tuning frequency range and (2) simulations do suggest that losses no greater that 5 dB and as low as ~1 dB are possible within the complete tuning range. Furthermore, these simulations provide encouragement that much lower losses are possible, even at frequencies as high as 20 GHz. Research in our group is focused on building devices with such performance levels. Performance characteristics of previously published MEMS RF filters are included in Table II for comparison. Our demonstrated approach is most competitive in regards to the tuning range. Given the simulations and room for optimization it is also believed that the approach is competitive with regards to the insertion loss. As it is a thermally driven actuator, while the drive voltage is low, the power consumption is high compared to capacitively tuned devices.

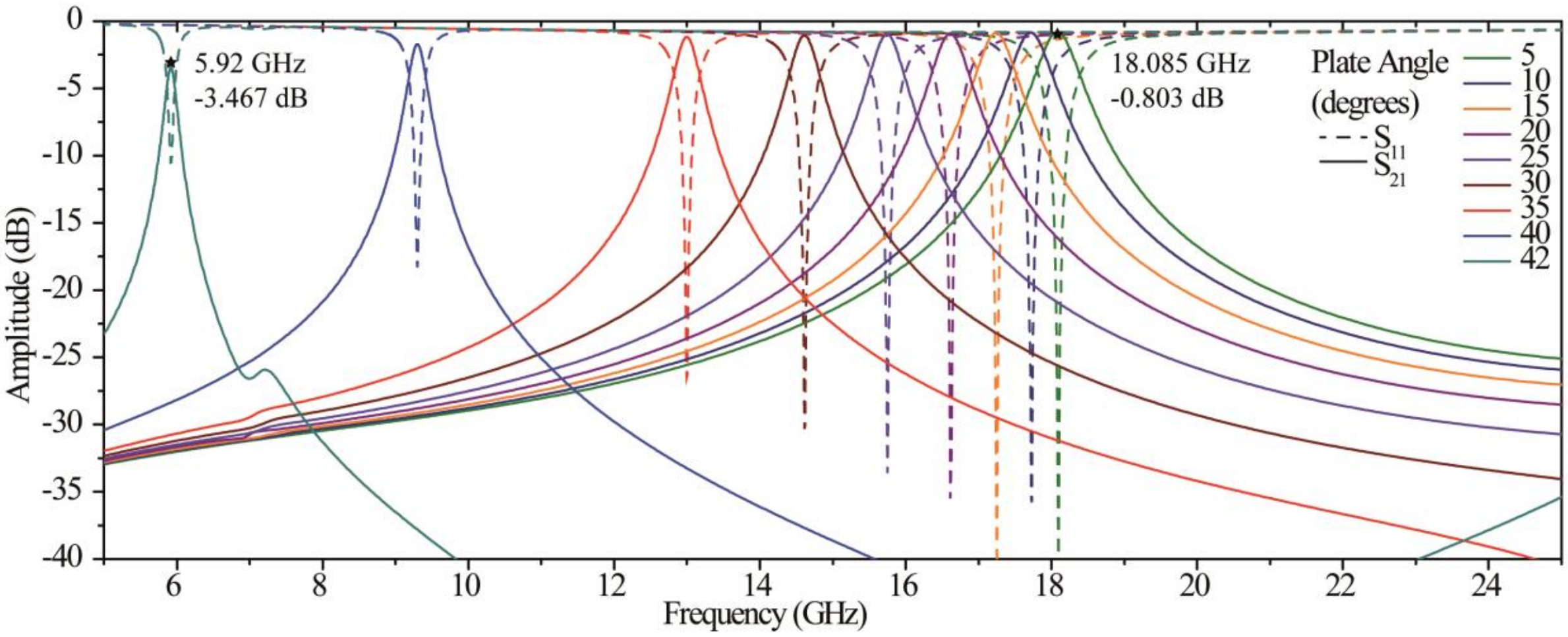

Figure 12. Shown are 3D EM simulations of our MEMS device and package. These simulations suggest very low losses are possible, perhaps approaching 1 dB.

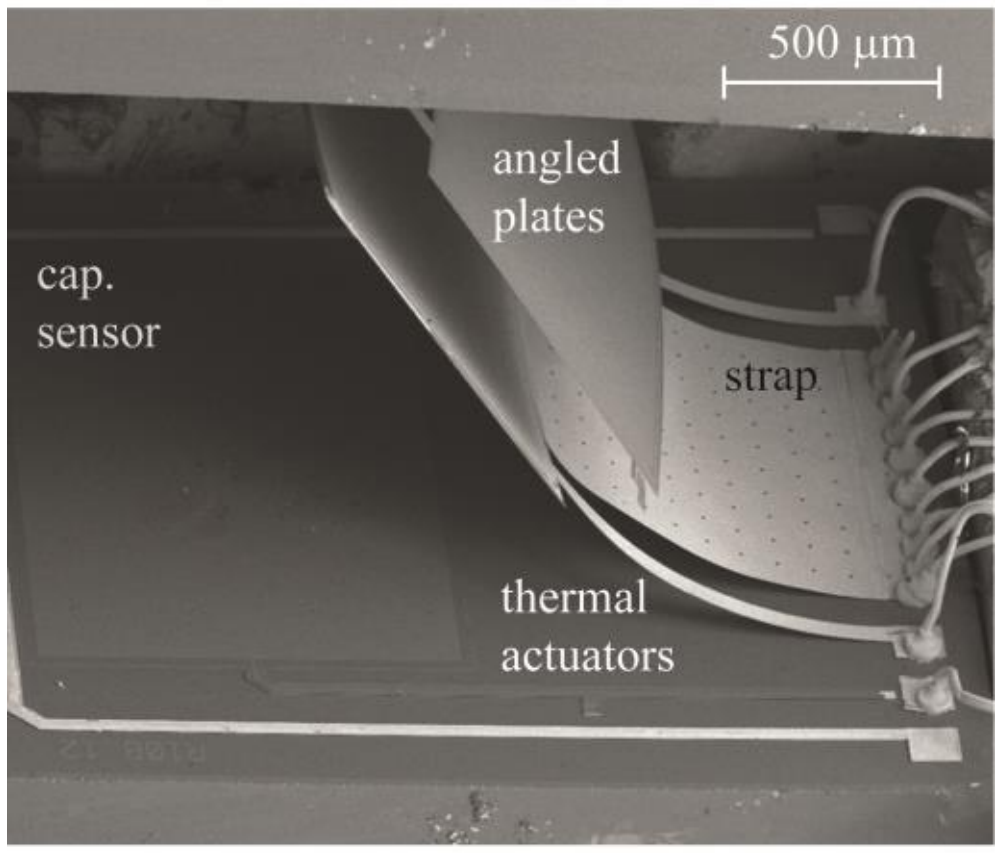

Figure 14. Shown is a two MEMS plate design. Rather than providing motion between a MEMS plate and a fixed post, this device works by tuning the distance between two MEMS plates, each attached to a die that are mounted facing each other. This version provides more intrinsic vibration isolation than does a single plate. One can just see the built-in fixed capacitor plate on the die substrate that is used to control the plate position.

## VII. CONCLUSIONS

In this paper we have described our approach to building high performance, MEMS microwave filters with analog tuning. To do this, we have developed a new class of surface micro-machined MEMS actuators that let us lift millimeter sized plates up to a millimeter from the surface. We have also developed a method of attaching a large gold strap to the MEMS plate, enabling low insertion losses and high *Q*. Our on-chip metrology for plate position and temperature allows us to precisely control the analog position of our plate and hence the center frequency of our filter. We have also described the small SWaP packaging enabled by our device and shown simulations suggesting < 1 dB losses are possible. Our best-case measured losses are ~2.5 dB. We believe our devices open up the possibility for small, fast low insertion loss microwave filters using a technology that could be integrated seamlessly with the other RF electronics. These devices may offer microwave engineers a compelling new type of filter for demanding RF systems in the GHz regime.

## ACKNOWLEDGEMENTS

DJB is supported by the Engineering Research Centers Program of the National Science Foundation under NSF Cooperative Agreement No. EEC-1647837.

TABLE II

SUMMARY OF PERFORMANCE PARAMETERS OF MEMS RF FILTERS. THE FIRST FOUR REFERENCES ARE MOST CLOSELY COMPARABLE TO THE DEVICES PRESENTED HERE. ADDITIONAL CONTINUOUS AND DISCRETE FILTER EXAMPLES ARE ALSO INCLUDED.

| Frequency (GHz) | 3 dB BW (%) | Tuning ratio (%) | Quality factor | Insertion Loss (dB) | Drive (V) | Continuous / discrete | Response time (μs) | Technology | Ref |
|---|---|---|---|---|---|---|---|---|---|
| 5.5-15 | 1.5-2.7 | 173 | 37-64 | 3-9 | 0-0.5 | continuous | ~1000 | 60° angled post and plate 3D mems thermally driven sweeping capacitance | this work |
| 20-40 | 1.9-4.7 | 100 | 264-540 | 0.8-2.9 | 0-175 | continuous | 10s | EVA MEMS | [18] |
| 4.07-5.58 | 0.8-1.1 | 37 | 354-400 | 3.18-4.91 | 60 | continuous | 25-50 | EVA MEMS | [19] |
| 18.6-21.44 | 7.35-7.67 (1dB) | 15 | 350-450 | 3.85-4.15 | 80 | discrete | <0.01 | MEMS bridges | [20] |
| 3.16-4.45 | 0.75 | 44 | 300-650 | 3.5-2.5 | 0-120 | continuous | 100s | MEMS 3D capacitive, comparable cavity | [9] |
| 1.56-2.48 | 7-9 | 63 | ~100 | 2.22-1.92 | 42 | discrete | 62 | MEMS capacitance tuning | [21] |
| 0.602-1.011 | 13-14 | 67 | 50 | 3-3.6 | 5 | discrete | 40-80 | MEMS single chip filter array | [22] |
| 14.73-15.34 | 3 | 4 | ~150 | 2.51-2.87 | | discrete (on/off) | | RF–MEMS Tunable Evanescent Mode | [23] |
| 2.35-3.21 | 2 | 37 | 356-405 | 1.65-1.42 | 110 | continuous | ~100 | EVA mode MEMS (capacitive) | [24] |
| 11 to 14 | 0.2 | 15 | 400-600 | 38-35 | 120 | continuous | | MEMS varactors and post capacitive | [25] |

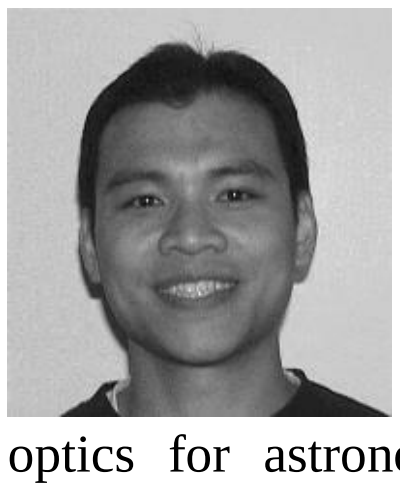

**Jackson Chang** received the M.Eng. in Materials Science and Engineering from Boston University in 2012. He currently works as a MEMS Design & Process Engineer at Boston Micromachines. His work focuses on MEMS deformable mirrors and their applications in adaptive optics for astronomy and asymmetric free space optical communication. Previously, he was a Research Engineer at the Solid State Research Lab at Boston University focusing on tunable plasmonic and RF MEMS technology.

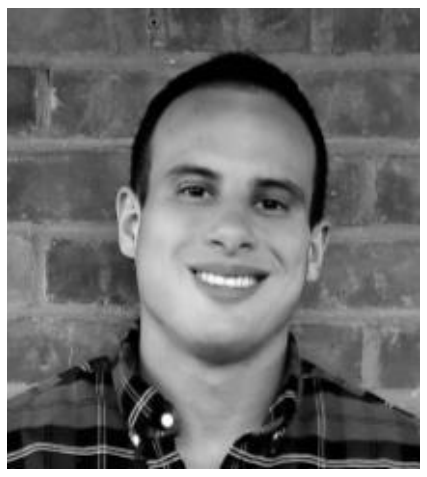

**Michael J. Holyoak** became a member of the IEEE in 2011. He received his BS in Engineering from The College of New Jersey in 2009 and his MS in Engineering from Columbia University in 2011. He is currently a Member of Technical Staff in the Wireless Solutions Group at LGS Innovations.

His work focuses on micro- and millimeter-wave radio systems, with emphasis in electromagnetics, antenna design, phased arrays, microelectronics packaging and assembly, and RF MEMS technology. Michael received the Industry Best Paper Award at the IEEE RFIC Symposium in 2015.

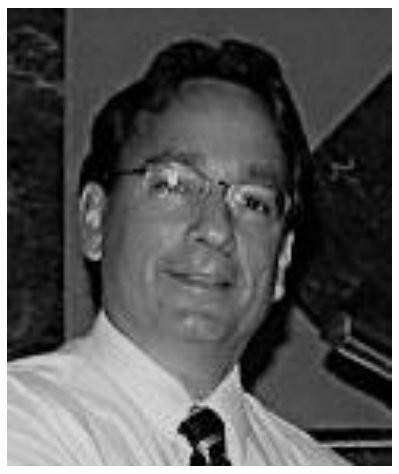

**George K. Kannell** received his MS degree from the New Jersey Institute of Technology in 1988.
George is the RF Systems Manager / Design Engineer at LGS Innovations where he is responsible for development of innovative communication products to over 100 GHz. At Bell Laboratories, he was responsible for System Architecture and realization of next generation Wireless Communications wireless systems. He has served as an adjunct Professor in Digital Communications at the New Jersey Institute of Technology. Prior to Bell Laboratories, he worked as a Senior Engineer at Ansoft Corporation in Simulation of Communications Systems, Microwave Circuits, RFIC circuits, Bipolar and FET device characterization and software development. Before this, at KDI/Triangle Electronics, he managed an engineering team in the design of Active and Passive Microwave Devices and Subsystems. He has published technical articles, holds several patents and presents at technical conferences. He serves as the Technical Program Chair for the IEEE Region 1. George is a member of Eta Kappa Nu and a Senior Member of the IEEE. George is a recipient of the IEEE Leadership Award and Bell Labs President's Awards.

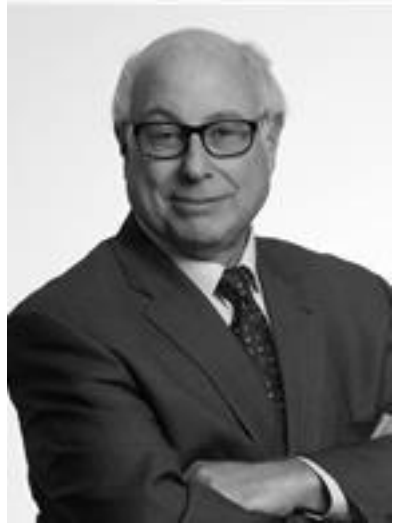

**Marc Beacken** became a member of the IEEE in 1978. He received his BS in Electrical Engineering from Rutgers College of Engineering, Rutgers University in 1978 and his MS in Electrical Engineering from Rensselaer Polytechnic Institute in 1980. He is currently a Senior Vice President (SVP) of LGS Innovations and Chief Technology Officer (CTO) of the Wireless Solutions business within LGS Innovations. Prior to LGS, Marc has had a long career in Bell Laboratories. He has received the distinction of Bell Labs Fellow and Distinguished Member of Technical Staff appointments. He has held technical and organizational positions of vice president, senior director, director, technical manager and principle investigator addressing both the federal and commercial marketplace providing innovative technology, platforms, products and end-to-end solutions in a diverse set of challenging and imimpactful application and technology areas. Areas of technical and publication activity include digital signal processing algorithms and systems; wide-band, multi-channel & tunable software defined radios; enabling component technology; multi-processing and multi-media systems; advanced network services platforms and applications; photonic processing; wireless communications & optical networking; ISR systems; and wireless technologies, data analytics and protocols.

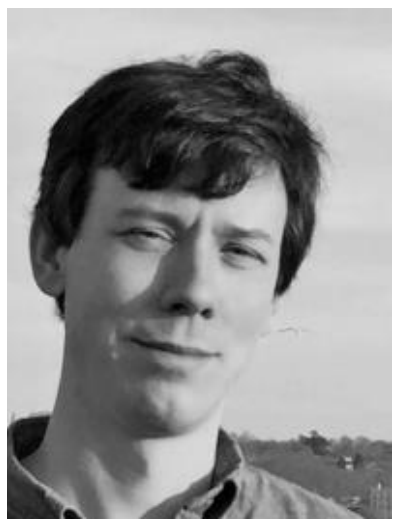

**Matthias Imboden** graduated from Bern University in 2004 with a Diploma (M.S.) in Physics. He obtained his physics Ph.D. titled *Diamond nano-electromechanical resonators: Dissipa-tion and superconductivity*, at Boston University in 2012.
He currently works as a post doctorate researcher at EPFL in the LMTS Lab. His work focuses on the study of MEMS devices for both research and technology development. Topics of interest include nanoscale fabrication, tunable plasmonics, smart lighting systems, and bioMEMS.
Dr Imboden is a member of the American Physics Society and IEEE.

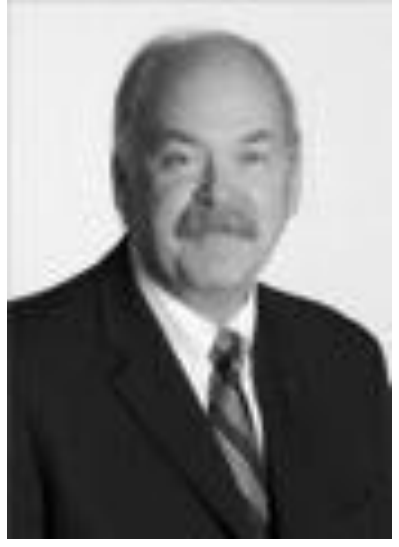

**David J. Bishop** became a member of the IEEE in 2011. He received his BS in Physics from Syracuse University in 1973 and his MS in Physics in 1977 and his Ph.D. in Physics in 1978, both from Cornell University.
He is currently the Head of the Division of Materials Science and Engineering, Boston University and also a Professor of Physics and a Professor of Electrical Engineering. Previously he was the Chief Technology Officer (CTO) and Chief Operating Officer (COO) of LGS, the wholly-owned subsidiary of Alcatel-Lucent dedicated to serving the U.S. federal government market with advanced R&D solutions. Before joining LGS, Dr. Bishop was the President of Government Research & Security Solutions for Bell Labs, Lucent Technologies. Dr. Bishop is a Bell Labs Fellow and in his previous positions with Lucent served as Nanotechnology Research VP for Bell Labs, Lucent Technologies; President of the New Jersey Nanotechnology Consortium and the Physical Sciences Research VP. He joined AT&T-Bell Laboratories Bell Labs in 1978 as a postdoctoral member of staff and in 1979 became a Member of the Technical Staff. In 1988 he was made a Distinguished Member of the Technical Staff and later that same year was promoted to Department Head, Bell Laboratories.

Professor Bishop is a member and fellow of the American Physical Society, a member of the MRS and a recipient of the APS Pake Prize.